\documentclass[aps,prl,twocolumn,showpacs,superscriptaddress,groupedaddress]{revtex4-1}
\usepackage{amsmath}
\usepackage{amssymb}
\usepackage{graphicx}% Include figure files
\usepackage{dcolumn}% Align table columns on decimal point
\usepackage{bm}% bold math
\usepackage{caption}
\usepackage{subcaption}
\usepackage{tikz}
\makeatletter
\newcommand\footnoteref[1]{\protected@xdef\@thefnmark{\ref{#1}}\@footnotemark}
\makeatother
\begin{document}

\preprint{APS/123-QED}

\title{A Microscopic Description of the Granular Fluidity Field in Nonlocal Flow Modeling}% Force line breaks with \\

\author{Qiong Zhang}
\author{Ken Kamrin}%
 \email{kkamrin@mit.edu}
\affiliation{%
Department of Mechanical Engineering, MIT, Cambridge, Massachusetts 02139, USA
}%

\date{\today}% It is always \today, today,

\begin{abstract}
A recent granular rheology based on an implicit  `granular fluidity' field has been shown to quantitatively predict many nonlocal phenomena. However, the physical nature of the field has not been identified. Here, the granular fluidity is  found to be a kinematic variable given by the velocity fluctuation and packing fraction. This is verified with many discrete element simulations, which show the operational fluidity definition, solutions of the fluidity model, and the proposed microscopic formula all agree. Kinetic theoretical and Eyring-like explanations shed insight into the obtained form.
\end{abstract}

\pacs{Valid PACS appear here}% PACS, the Physics and Astronomy
\maketitle
%%%%%%%%%%%%%%%%%%%%%%%%%%%%%%%%%%%%%%%%%%%%%%%%%%%%%%%%%%%%%%%%%%%%%
%review
The rheology of dry granular materials is commonly studied in homogeneous simple (planar) shear tests. In these tests, the \emph{inertial granular rheology} can be observed, in which a one-to-one relationship exists between two dimensionless numbers: $\mu=\tau/P$, the ratio of shear stress $\tau$ to normal stress $P$, and the inertial number $I=\dot{\gamma}d/\sqrt{P/\rho_s}$, which nondimensionalizes the shear rate $\dot{\gamma}$ by the (mean) particle size $d$, $P$, and the solid density $\rho_s$ \cite{andreotti2013granular, da2005rheophysics}.  Empirically, the bijection between $\mu$ and $I$ is often fitted to the form \cite{jop2005crucial}
\begin{align}
    \mu=\mu_{loc}(I)=\mu_s+\frac{\Delta \mu}{I_0/I+1},
    \label{local_law}
\end{align}
where $\mu_s$ is a static yield value, below which the system does not flow, $\mu_2$ is an upper limit for $\mu$ at high rates, $\Delta \mu=\mu_2-\mu_s$, and $I_0$ is a dimensionless constant.
%%%%%%%%%%%%%%%%%%%%%%%%%%%%%%%%%%%
% nonlocal effect review
% review of other views or models
Despite its effectiveness in steady simple shearing, granular behavior in more general circumstances can be observed to deviate from the inertial rheology. In inclined plane flows, where the $\mu$ field is spatially homogeneous and given by the tilt angle, the angle at which a flowing layer stops depends explicitly on the size (thickness) of the pile \cite{silbert2003granular,midi2004dense}. In steady but non-uniform flow geometries, flow is observed in zones where $\mu<\mu_s$ and the $\mu-I$ relation is not one-to-one in these regions \cite{da2002viscosity,koval2009annular}. Moreover, a ``secondary rheology'' has been observed in which the dynamics of a loaded probe submerged in quiescent material is influenced by the motion of far-away boundaries of the granular system \cite{reddy2011evidence,nichol2010}. Such phenomena deviating from the inertial law are describable only by considering nonlocal effects. Various microscopic notions have been considered to understand the origins of this nonlocality \cite{radjai2002turbulentlike,pouliquen2004velocity,lois2006emergence,staron2008correlated,staron2010flow,bouzid2013nonlocal}.

%%%%%%%%%%%%%%%%%%%%%%%%%%%%%%%%%%%
Recently, a size-dependent granular rheological framework has been proposed based on a state field called the ``granular fluidity.'' With minimal fitting parameters, the model has shown the capability of quantitatively predicting a range of nonlocal effects in multiple geometries, including all the deviations from $\mu(I)$ behavior described above \cite{kamrin2012nonlocal,henann2013predictive,kamrin2014effect,henann2014thermomechanics,henann2014secondary,kamrin2015nonlocal}.  The granular fluidity field, denoted $g$, is presumed to be governed by a dynamical partial differential equation (PDE) \cite{kamrin2015nonlocal}:
\begin{align}
    t_0\dot{g}=A^2d^2\nabla ^2g-\Delta \mu \left( \frac{\mu_s-\mu}{\mu_2-\mu} \right)g -b\sqrt{\frac{\rho_s d^2}{P}}\mu g^2
    \label{g_pde}
\end{align}
where $A$ is a dimensionless constant called the \emph{nonlocal amplitude}, $t_0$ is a time-scale, and $b=\Delta \mu/I_0$ \footnote{This the full form of the granular fluidity relation; an approximation for steady-state solutions only is also commonly used \cite{kamrin2012nonlocal,henann2013predictive}.}. The $g$ field influences the flow through its role in the constitutive relation between stress and strain-rate: $\dot{\gamma}=g\mu$. Together, the result is a flow model with an intrinsic length-scale given by $d$. The inertial law, Eq.~(\ref{local_law}), can be obtained when the flow field is homogeneous ($\nabla g=\vec{0}$) and in steady state. While these equations define the model from a mathematical perspective, the physical nature of the granular fluidity field is not clear. To be valid in its role within the constitutive model, granular fluidity should be a kinematically observable state variable, as was stressed in \cite{bouzid2015non} and was assumed in \cite{henann2014thermomechanics} where Eq (\ref{g_pde}) was reconciled form a variational argument.    What \emph{is} the granular fluidity?

%%%%%%%%%%%%%%%%%%%%%%%%%%%%%%%%%%%%%%%%%%%%%%%%%%%%%%%%%%%%%%%%%%%%%
In this letter, we identify a microphysical definition for the granular fluidity field, which defines the fluidity in terms of the velocity fluctuation and packing fraction, two kinematic variables. Using discrete element method (DEM) simulations in multiple configurations at steady-state, we compare the predictions of this microscopic formula with the constitutive definition $g= \dot{\gamma}/\mu$, as well as steady solutions of $g$ from the PDE for granular fluidity.  The strong agreement found gives evidence that the PDE is in fact a model for the behavior of this kinematic field. Lastly, we attempt to explain the microphysical description of $g$ using kinetic theory and also using an Eyring-like model, which illustrates a possible fluctuation activated process of granular flows.

%%%%%%%%%%%%%%%%%%%%%%%%%%%%%%%%%%%%%%%%%%%%%%%%%%%%%%%%%%%%%%%%%%%%%
We start with the hypothesis that for hard particles, $g$ should relate to velocity fluctuations \cite{bocquet2001granular,forterre2001longitudinal,utter2008experimental,artoni2015effective,reddy2011evidence}, $\delta v$, among other possible state variables. From its operational usage in the flow-rule, $g=\dot{\gamma}/\mu$ has dimensions of inverse time.  We propose the relevant time scale is $d/\delta v$ such that the fluidity $g$ can be nondimensionalized as $gd/\delta v$. If we suppose the only other state variable affecting $g$ is the packing fraction $\Phi$, the normalized fluidity should be expressible as
\begin{align}
    \frac{gd}{\delta v}=F(\Phi) \Rightarrow
    g=\frac{\delta v}{d}\cdot F(\Phi),
    \label{g_form}
\end{align}
where $F$ is an unknown function.

%%%%%%%%%%%%%%%%%%%%%%%%%%%%%%%%%%%%%%%%%%%%%%%%%%%%%%%%%%%%%%%%%%%%%
% description of the simulation
To evaluate the hypothesis, DEM simulations of three kinds of configurations were implemented in the open-source software LAMMPS \cite{plimpton1995fast}. The particles simulated are spheres with solid density $\rho_s=2500 \mbox{kg/m}^3$, mean diameter $d=0.0008 \mbox{m}$, and polydispersity of 20\% to prevent crystallization. The particle interaction model \cite{koval2009annular,kamrin2014effect} contains elastic forces, damping effects, and Coulomb friction using a spring-dashpot law defined by stiffness in the normal and tangential directions $k_n$, $k_t=2/7 k_n$, damping coefficient in the normal and tangential directions $\gamma_n$, $\gamma_t$ and surface friction coefficient $\mu_c=0.4$. We adopt $\gamma_t$ to be 0 and calculate $\gamma_n$ with the restitution coefficient $e=0.1$: $\gamma_n=-2 \text{ln}e\sqrt{mk_n/(\pi^2+\text{ln}^2e)}$. Throughout, $k_n/Pd^2 > 10^4$ is kept, which ensures that the deformation of the particles is small enough to be in the hard particle regime. The simulated system is a cuboid domain ($20d\times8d\times\sim50d$, 9096 particles in total) of particles sheared between two planar rough walls made of particles of the same properties at the top and bottom. The bottom wall is fixed in all the cases. If a pressure boundary condition at the top wall is needed, the top wall's height $h$ is controlled to set the pressure using a feedback process \cite{koval2009annular}. To apply a fixed volume boundary condition, the top wall is held stationary in $z$. Periodic boundary conditions are applied to the other four boundaries. The time step is chosen as $dt=\sqrt{m/(50k_n)}$.

\begin{figure}
    \begin{subfigure}{0.135\textwidth}
        \includegraphics[height=6cm]{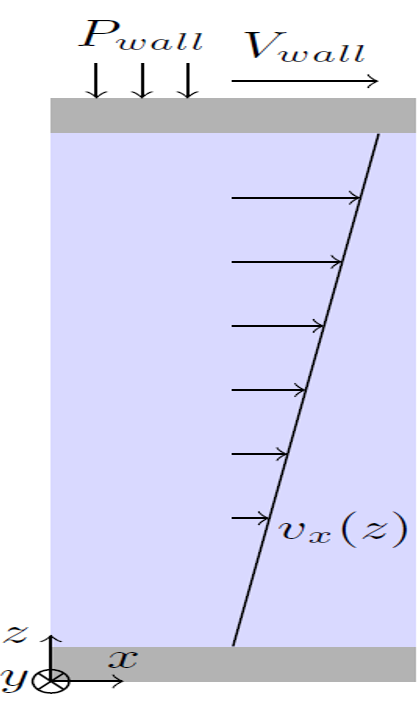}
        \caption{}
        \label{fig1: geometries(a)}
    \end{subfigure}
    \begin{subfigure}{0.135\textwidth}
        \includegraphics[height=6cm]{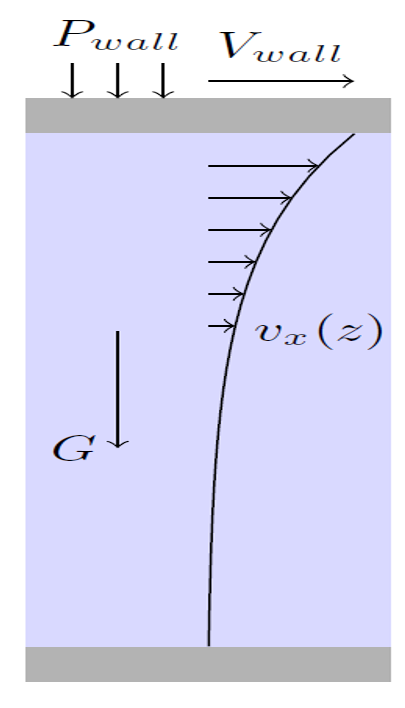}
        \caption{}
        \label{fig1: geometries(b)}
    \end{subfigure}
    \begin{subfigure}{0.135\textwidth}
        \includegraphics[height=6cm]{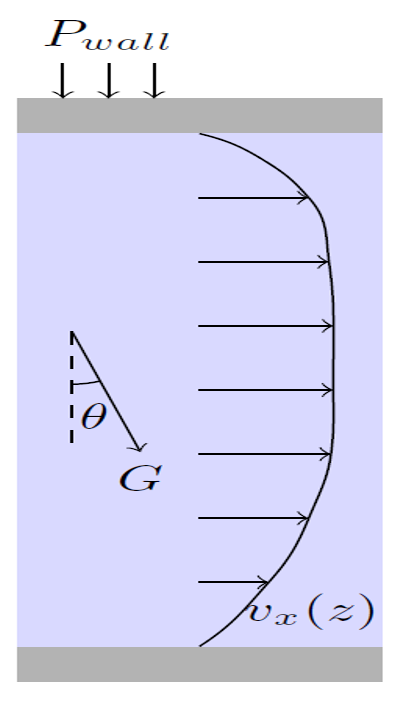}
        \caption{}
        \label{fig1: geometries(c)}
    \end{subfigure}
    \captionsetup{justification=raggedright,
singlelinecheck=false
}
    \caption{Configurations of granular flow geometries tested, with qualitative velocity profiles plotted. (a) planar shear; (b) planar shear with gravity; (c) chute flows. In chute flow cases with fixed volume boundary conditions instead of pressure control (not pictured) the top wall is stationary.}
    \label{fig1:geometries}
\end{figure}

When the steady state is reached, we output data every 20,000 steps, collecting a total of $N$=3000 snapshots for each simulation.
We used three families of configurations: homogeneous planar shear, planar shear with gravity, and chute flows, as shown in FIG.~\ref{fig1:geometries}. The confining pressure at the top boundary is $P_{wall}=P_f\cdot P_0$, where $P_0=5\times10^{-6}k_n/d$, the gravity is $G=G_f\cdot g_0$, where $g_0=0.1P_0d^2/m$, and the horizontal velocity of the top wall is $V_{wall}=V_f\cdot \sqrt{6P_0/\pi \rho_s}$, where $P_f$, $G_f$ and $V_f$ are dimensionless factors to control the boundary conditions and the gravity. In gravity-free planar shear cases, the confining pressure is chosen as $P_f=1$ and ten different $V_f$'s are examined. In the cases of planar shear with gravity, $G_f$ is fixed at $G_f=1$ and five cases are simulated with four different $V_f$'s and two $P_f$'s. In chute flow cases, inclined angles of $\theta=90^\circ,75^\circ,60^\circ,45^\circ$ are simulated. Gravity is increased to $G_f=3$ in the slanted configurations and the top wall is fixed in the $x$ direction, $V_f=0$. We test fixed volume top-wall constraints for all chute cases. We also perform fixed top-wall pressure constraints in the  $\theta=90^\circ$ chute case, where, to make a direct comparison, the confining pressure is chosen to be the same as the mean pressure at the top wall in the corresponding fixed volume case. In total, twenty different flows are simulated.
%%%%%%%%%%%%%%%%%%%%%%%%%%%%%%%%%%%%%%%%%%%%%%%%%%%%%%%%%%%%%%%%%%%%%
% averaging method and definition of variables

In all geometries, the time averaged fields should be homogeneous in each horizontal ($x-y$) plane. Dividing the simulated domain into layers at differing heights $z$, we first take the spatial average of the variable of interest in the whole layer ($x-y$ planes), and then average the instantaneous layer-wise values arithmetically in time.
%%%%%%% velocity fluctuation
When the layer-wise instantaneous velocity has been calculated, we subtract the instantaneous mean velocity from the particle velocity to obtain the velocity deviation. The velocity fluctuation $\delta v$ is defined as the root of \emph{granular temperature} which is the mean square of velocity deviation.
%%%%%%% stress tensor
The particle-wise stress tensor is calculated considering both the contact contribution and the kinetic contribution, as defined in \cite{koval2009annular}. After the average Cauchy stress tensor has been obtained in a layer, the pressure is defined as $P=-\sigma_{ii}/3$ and the shear stress is defined as the equivalent shear stress $\tau=\sqrt{\sigma'_{ij}\sigma'_{ij}/2}$, where $\sigma'_{ij}=\sigma_{ij}+P\delta_{ij}$ is the stress deviator. See Supplemental Material \footnote{See Supplemental Material for more details of the averaging methodology, verifications of the averaging method, detailed comparisons with kinetic theory, and further discussion on whether $g=g(\delta v, \Phi)$ can be reduced\label{note1}} for the detailed averaging method, its verification, and particular details of the calculation of $\delta v$.

To be free of wall effects, we have excluded layers at distances smaller than $4d$ from the walls. To ensure that the variables are at steady-state, we accept data from layers that have experienced a strain no smaller than $25$ in sampling. The span of inertial numbers in the data presented is 0.0007 to 0.6, which covers the range of flow regimes, from quasi-static to collisional.

%%%%%%%%%%%%%%%%%%%%%%%%%%%%%%%%%%%%%%%%%%%%%%%%%%%%%%%%%%%%%%%%%%%%%
% test the hypothesis and make prediction
\begin{figure}[h]
    \includegraphics[trim = 0mm 0mm 8.5mm 6mm, width=0.4\textwidth]{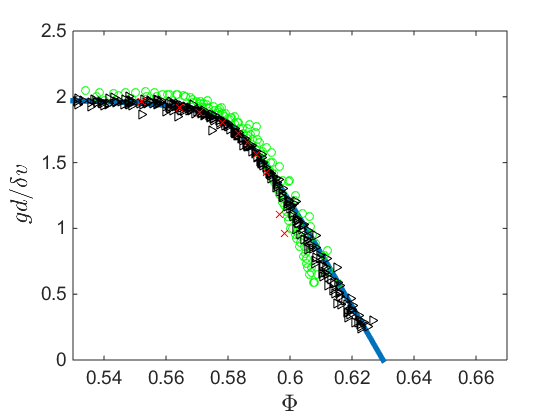}
  \captionsetup{justification=raggedright, singlelinecheck=false}
  \caption{Normalized fluidity plotted against packing fraction. Data from all chute flow tests ($\rhd$), planar shear with gravity tests ($\circ$) and homogeneous planar shear tests ($\times$). A hyperbola (solid line) is fit for $F(\Phi)$.}
  \label{fig3:verification}
\end{figure}

To determine if $g=F(\Phi)\delta v/d$, we calculate the layer-wise fluidity field $g=\dot{\gamma}/\mu$,  $\delta v$, and $\Phi$ in all cases. For added precision, here we evaluate $d$ as the layer-wise mean particle size; though quasi-monodisperse, the packing shows slight ($\lesssim 10\%$) spatial variation in $d$. We plot normalized fluidity $gd/\delta v$ against packing fraction $\Phi$ in FIG.~\ref{fig3:verification} for all 20 flows simulated.   Each data point corresponds to a different $z$ value, except gravity-free planar shear cases, which are nearly uniform and show a single data point per test. $\Phi$ is smoothed in a band as wide as $\pm 2d$ at $z$ intervals of $0.5d$.
The data from different configurations collapse well, suggesting that packing fraction and velocity fluctuation are sufficient to define the granular fluidity. $F(\Phi)$ has a nearly constant behavior for low $\Phi$ values, which transitions into a roughly linear decrease at high values. This behavior can be fit to a hyperbola $F(\Phi)=\frac{-(\Phi-0.58)+\sqrt{(\Phi-0.58)^2+1.54\times 10^{-4}}}{0.048}+2.0,$ which vanishes at $\Phi=0.63$, approximately random close packing. An interpretation for the functional form is given later.

\begin{figure*}
\begin{subfigure}{0.245\textwidth}
\includegraphics[trim = 0mm 0mm 8.5mm 6mm, clip,width=\linewidth]{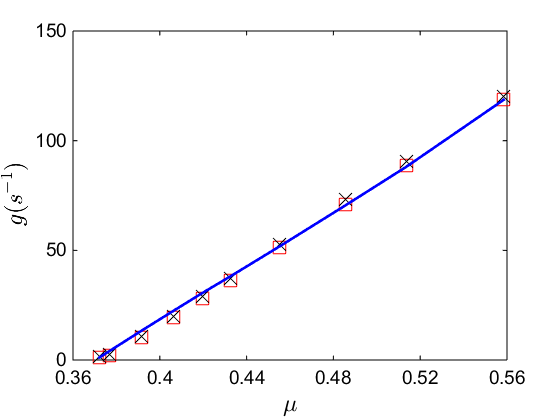}
\end{subfigure}
\begin{subfigure}{0.245\textwidth}
\includegraphics[trim = 0mm 2mm 8.5mm 8mm, clip,width=\linewidth]{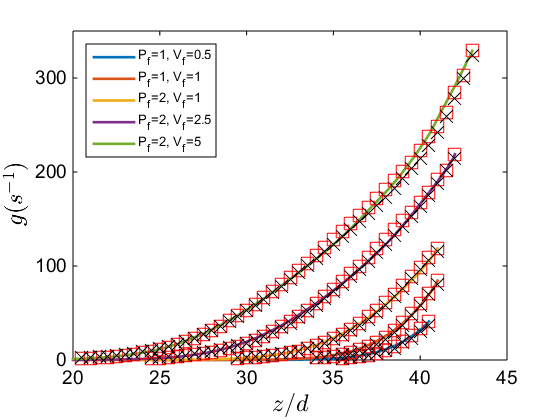}
\end{subfigure}
\begin{subfigure}{0.245\textwidth}
\includegraphics[trim = 0mm 2mm 10mm 5mm, clip,width=\linewidth]{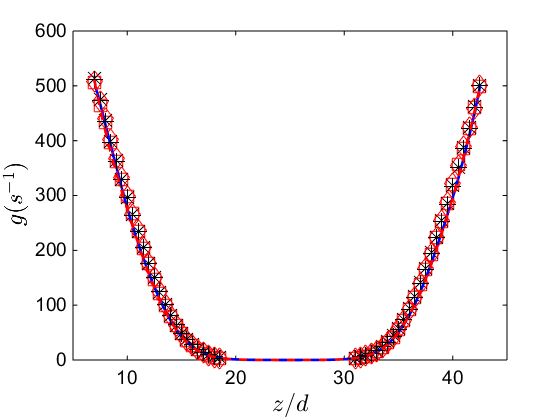}
\end{subfigure}
\begin{subfigure}{0.245\textwidth}
\includegraphics[trim = 0mm 2mm 10mm 5mm, clip,width=\linewidth]{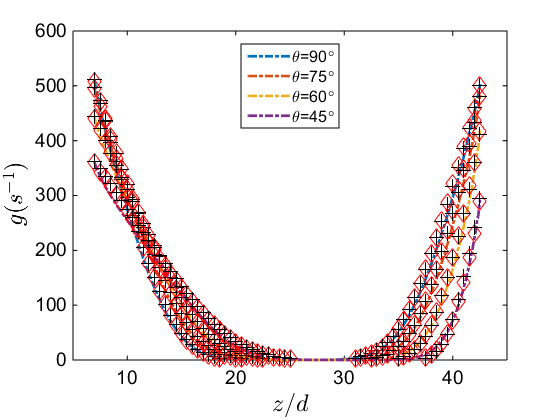}
\end{subfigure}
\hfill\\
\begin{subfigure}{0.245\textwidth}
\includegraphics[trim = 0mm 0mm 8.5mm 6mm, clip,width=\linewidth]{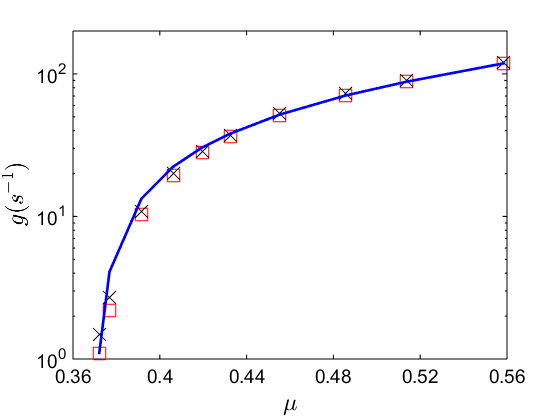}
\caption{\label{fig4: predictions(a)}}
\end{subfigure}
\begin{subfigure}{0.245\textwidth}
\includegraphics[trim = 0mm 2mm 8.5mm 8mm, clip,width=\linewidth]{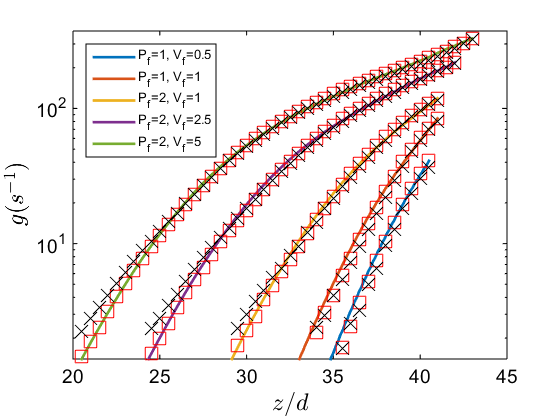}
\caption{\label{fig4: predictions(b)}}
\end{subfigure}
\begin{subfigure}{0.245\textwidth}
\includegraphics[trim = 0mm 2mm 10mm 5mm, clip,width=\linewidth]{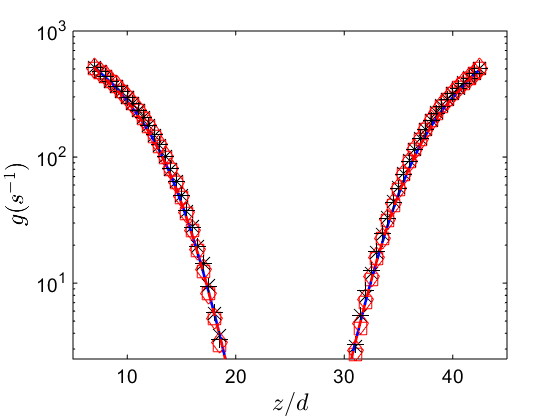}
\caption{\label{fig4: predictions(c)}}
\end{subfigure}
\begin{subfigure}{0.245\textwidth}
\includegraphics[trim = 0mm 2mm 10mm 5mm, clip,width=\linewidth]{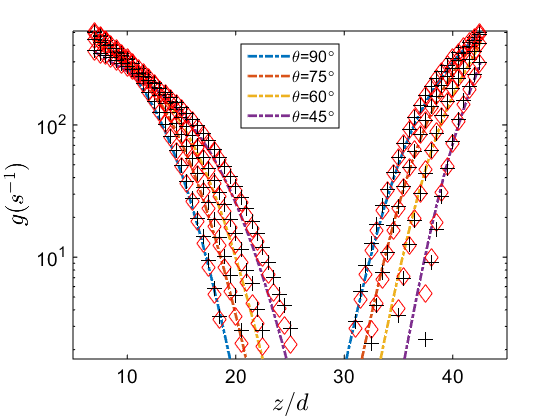}
\caption{\label{fig4: predictions(d)}}
\end{subfigure}
\captionsetup{justification=raggedright,
singlelinecheck=false
}
\caption{Comparison of three definitions of $g$: DEM results of $\dot{\gamma}/\mu$ ($\square, \diamond$), solutions of Eq.~(\ref{g_pde}) (lines), and predictions of the microscopic formula, Eq.~(\ref{g_form}) ($\times,+$). Comparisons in  (a)  homogeneous planar shear cases, (b) planar shear with gravity, (c) $90^{\circ}$ chute flow cases with different boundary conditions, and (d) chute flow cases at different inclinations.  Results of DEM simulations using fixed volume BC's are indicated with ($\diamond, +$) and fixed wall pressure BC's with ($\square, \times$).}
\label{fig4: predictions}
\end{figure*}

For further examination, in FIG.~\ref{fig4: predictions} the predicted fluidity fields using the microscopic formula are compared with $\dot{\gamma}/\mu$ and solutions of Eq.~(\ref{g_pde}) in all configurations tested. In the PDE, $b=1.041$, $\mu_s=0.3704$ and $\mu_2=0.95$ are obtained by fitting the $\mu(I)$ relation in the homogeneous planar shear cases as shown in FIG.~\ref{fig4: predictions(a)}, and we choose $A=0.44$.  All these values are close to the values found for glass spherical beads \cite{jop2005crucial,henann2013predictive}, with the exception of $\mu_2$ which we find to be larger in our discrete simulations \footnote{The value of $t_0$ is irrelevant here, as we study only steady-state.}. The boundary condition for $g$ is taken as the wall value of $\dot{\gamma}/\mu$ from DEM and the PDE is run to steady-state using a finite-difference method to obtain steady $g$ profiles. Note, the fluidity profiles of $90^\circ$ chute flow cases with fixed volume boundary conditions and fixed pressure boundary conditions, compared in FIG.~\ref{fig4: predictions(c)}, match well, showing that the type of boundary condition does not have a significant influence on the constitutive behavior in the interior. This is also the case if the incline angle is varied, which we verified in additional tests.
Overall, the collapse of the three different definitions of $g$ in the various cases is strong, evidencing the generality of the $g=F(\Phi)\delta v/d$ formula.

%%%%%%%%%%%%%%%%%%%%%%%%%%%%%%%%%%%%%%%%%%%%%%%%%%%%%%%%%%%%%%%%%%%%%
% interpretation in kinetic theory and Eyring equation

There are at least two ways to reconcile these results physically, one with kinetic theory and another in terms of an activated process. Regarding kinetic theory, in Lun et al.'s work \cite{lun1984kinetic}, $P$ and the viscosity $\eta$ in a granular gas are functions of $\Phi$ and granular temperature $T$:
\begin{align*}
    &P(\Phi,T)=\rho F_1(\Phi)T &\text{and} 
    &&\eta(\Phi,T)=\rho dF_2(\Phi)\sqrt{T},
\end{align*}
where $\rho$ is the density $F_1(\Phi)$ and $F_2(\Phi)$ are functions of packing fraction $\Phi$. Under these relations, the operational definition of fluidity would imply
\begin{align*}
    g=\frac{\dot{\gamma}}{\mu}
     =\frac{P}{\eta}
     =\frac{\rho F_1(\Phi)T}{\rho dF_2(\Phi)\sqrt{T}}
     =\frac{\sqrt{T}}{d}\frac{F_1(\Phi)}{F_2(\Phi)}
     =\frac{\delta v}{d}\frac{F_1(\Phi)}{F_2(\Phi)},
\end{align*}
whose form is the same as Eq.~(\ref{g_form}). Though the fitted $F(\Phi)$ is different from $F_1(\Phi)/F_2(\Phi)$ given in \cite{lun1984kinetic}, the similarity of the form is suggestive.
Also, since the nonlocal effects captured by the fluidity model are most evident in quasi-static regions with enduring contacts, it seems the extended kinetic theory \cite{jenkins2010dense} would be needed to further this connection, in which dense behavior beyond binary collisions is modeled.  Kinetic theory based justifications have previously been used to explain the apparent density- and temperature- dependent viscosity of dense granular flows in annular shear experiments \cite{losert2000particle}, which foreshadows the functional form shown above. See Supplemental Materials \footnotemark[1] for more analysis of the connection with these theories.

Eyring's equation \cite{eyring1936viscosity} for activated processes has an analogy in the flow behavior of amorphous solids \cite{spaepen1977micro} and has been considered previously in the context of granular flows \cite{pouliquen2009non,reddy2011evidence}.  Here, as in \cite{spaepen1977micro}, we can view a material volume element as a collection of microscopic `sites'. Each site able to undergo a shear event or `hop' has a barrier that must be overcome in order to do so. We express the element's total shear rate through the product:
\begin{align}
       & \dot{\gamma}=\left(    \begin{array}{cc}
         & \text{Fraction} \\
         & \text{of sites}\\
         & \text{able to}\\
         & \text{shear}
    \end{array}         \right)
    %\cdot
    \left(\begin{array}{cc}
         & \text{Net \# of} \\
         & \text{positive shear} \\
         & \text{events per site} \\
         & \text{per second}
      \end{array}         \right)
        \left(    \begin{array}{cc}
         & \text{Strain} \\
         & \text{per} \\
         & \text{shear} \\
         & \text{event} \\
      \end{array}         \right)
    \label{Eyring_like}
\end{align}
We assume \cite{spaepen1977micro} that the fraction of potential sites for a shear event is a function of packing fraction, $f_1(\Phi)$. It should approach $0$ at a jammed $\Phi$ value, $\approx 63\%$. The second term can be expressed as the product of an attempt frequency for perturbations, $\omega$, and the net probability, $Pr^+-Pr^-$, that a perturbation causes a positive shear event vs a backward (negative) event on a site. The frequency of attempts should be related to the fluctuational motion of the particles, so we assume  that $\omega$ is given by $\delta v\ f_2(\Phi)/d$, i.e. the ratio of fluctuation velocity to a characteristic length $d/f_2(\Phi)$. The exponent of the probability of reaction is typically given by an Arrhenius form, which depends on an energy barrier and a distribution of energies in an attempt, which depends on temperature. The limit of rigid particle interactions in granular systems complicate energetic considerations in this context, but one can instead express $Pr^{\pm}$ in terms of a critical shear stress barrier $\Delta^{\pm}$ for forward or reverse shear events on a site, and a probability distribution for the stress fluctuation delivered by a perturbation, as in \cite{pouliquen2009non}. We similarly assume that $\Delta^{\pm}=\mu_2 P\mp\tau$, where the mean applied shear stress on the element, $\tau$, is seen to bias the barrier, and that the cumulative distribution function for stress perturbations is exponentially decaying, as observed in \cite{radjai1999contact,mueth1998force,majmudar2005contact}, with mean on the order of  $P$.  Therefore,
\begin{align*}
    Pr^+-Pr^-=C\left(\text{e}^{-\frac{\mu_2P-\tau}{P}}-\text{e}^{-\frac{\mu_2P+\tau}{P}}\right)=2C\text{e}^{-\mu_2} \text{sinh}\mu
\end{align*}
where $C$ is a constant factor.  Lastly, we make the common assumption \cite{spaepen1977micro,kamrin2014two,pouliquen2009non} that the strain per flow event is a constant, $\gamma_0$, related to a typical `jump distance'. We can now multiply the three factors together and approximate $\mu \approx \sinh \mu$ to get
$
    \dot{\gamma}
    \approx \gamma_0 \frac{\delta v }{d}f_1(\Phi)f_2(\Phi) 2C \text{e}^{-\mu_2}\mu.
$
Because $\mu<\mu_2=0.95$, the error of the linear approximation is always less than $15\%$, and much smaller in quasi-static media.  Upon dividing by $\mu$ one obtains
\begin{align}
    g=\frac{\dot{\gamma}}{\mu}
    =\frac{\delta v }{d}\cdot \underbrace{2C \gamma_0 \text{e}^{-\mu_2}f_1(\Phi)f_2(\Phi)}_{F(\Phi)} .
    \label{Eyring_explain}
\end{align}
Since $\mu_2$, $C$, and $\gamma_0$ are constant, Eq.~(\ref{Eyring_explain}) has the same form as Eq.~(\ref{g_form}). This analysis suggests shear flow in granular media may be a fluctuation activated process. 

Supposing $f_1$ is the dominant contribution to $F$, FIG.~\ref{fig3:verification} suggests a natural interpretation.  When $\Phi<0.57$, the packing is open enough that all sites are able to flow, and $f_1$ holds at its maximum value in this range.  As packing fraction increases above $0.57$, the number of sites that have enough free volume for flow shows an expected monotonic decrease. The Cohen-Turnbull theory of free-volume distribution in glassy materials gives a similar behavior for the fraction of sites above a critical free-volume threshold \cite{cohen1959molecular}.

Since Eq.~(\ref{g_pde}) evolves $g$ as a single state variable, it is natural to ask if the observed relation $g(\delta v, \Phi)$ can be reduced to depend on only one state variable rather than two.  For example, through the assumption of a local equation of state, one might suppose the pressure could be used to eliminate $\Phi$ or $\delta v$ from the system. In view of the data from the many geometries tested, we find this is not the case (see Supplemental Material \footnotemark[1]), suggesting the equation of state in inhomogeneous dense flow may need to include dependence on gradients and/or nonlocal influences. Because $\Phi$ and $\delta v$ both play an irreducible and influential role in defining $g$, a microphysical derivation for the granular fluidity PDE, Eq.~(\ref{g_pde}), may involve combining dominant terms from a heat equation and a density evolution rule, though first attempts using kinetic theory forms produce a system more complex than Eq.~(\ref{g_pde}). Deriving Eq.~(\ref{g_pde}) from the microscopic description of $g$ remains crucial future work.

%%%%%%%%%%%%%%%%%%%%%%%%%%%%%%%%%%%%%%%%%%%%%%%%%%%%%%%%%%%%%%%%%%%%%
Herein, we have proposed a relation connecting granular fluidity to granular velocity fluctuation and packing fraction. This relation is demonstrated in DEM simulations of multiple configurations. All three descriptions of the granular fluidity field --- (i) its operational definition ($g=\dot{\gamma}/\mu$) extracted from DEM simulations, (ii) its definition from the fluidity governing PDE, and (iii) the new microphysical definition --- match well with each other in multiple geometries under multiple conditions. Granular fluidity could be related to a fluctuation activated process as a measure of the rate of the number of flowable microsites perturbed by attempts. It is also interesting to compare the description herein to previous models for non-granular materials (e.g. emulsions, suspensions), which evolve the `standard' fluidity $f=\dot{\gamma}/\tau$ (inverse viscosity) rather than $g$. Theories have previously suggested $f$ relates microscopically to the rate of plastic events \cite{bocquet2009kinetic}, which has been correlated to shear-rate fluctuations in experiments \cite{jop2012microscale}.  Our result that a fluctuation variable is key as well in describing granular fluidity suggests a bridge between the different amorphous material classes.

%%%%%%%%%%%%%%%%%%%%%%%%%%%%%%%%%%%%%%%%%%%%%%%%%%%%%%%%%%%%%%%%%%%%%%%%%%%%%%%%%%%%%%%%%%%%%%
\section*{Supplemental material}
\subsection*{The averaging method and the definition of velocity fluctuation}
In all geometries, the time averaged fields should be homogeneous in each horizontal ($x-y$) plane. The simulated domain is divided into 100 horizontal layers at differing heights $z$, at intervals of $0.5d$. Throughout, to calculate layer-wise variables, we first take the spatial average of some variable of interest, say $\zeta$, in the whole layer ($x-y$ planes) and then average the instantaneous values, $\bar{\zeta}(z=z_k,t=t_i)$, arithmetically in time, $\bar{\zeta}(z=z_k)=\frac{1}{N}\sum_{i=1}^{N} \bar{\zeta}(z=z_k,t=t_i)$. For physically consistent spatial averaging, kinematic variables are weighted by the areas of the cross-sections of the particles in the corresponding layers. For example, the instantaneous spatially averaged velocity in layer $z=z_k$ is $\overline{\vec{v}}(z=z_k,t_i)= \left(\sum_{j}A_{j,i,k}\cdot \vec{v}(\vec{x}_j,t_i)\right)/\left(\sum_{j}A_{j,i,k}\right),$ where $\vec{v}(\vec{x}_j,t_i)=\vec{v}_{j,i}$ is the particle-wise velocity at position $\vec{x}_j$ and time $t_i$ and $A_{j,i,k}$ is the area of cross-section of $j$th particle in the layer $z=z_k$ at time $t_i$. The layer-wise mean velocity fluctuation $\delta v$ is calculated differently as the square root of \emph{granular temperature}:
\begin{align*}
    \overline{\delta v}^2 (z=z_k,t=t_i)=\frac{\sum_{j}A_{j,i,k} \left( \vec{v}(\vec{x}_j,t_i) - \overline{\vec{v}}(z=z_j,t_i) \right)^2 }{\sum_{j}A_{j,i,k}}.
\end{align*}
As defined above, before calculating the velocity fluctuation, we need the instantaneous mean velocities of each layer and then interpolate the particle-wise mean instantaneous velocity using the position $\vec{x}_j$ of each particle between two layers. Then the squared velocity deviations are averaged in both space (shown above) and time. This definition is quite similar to the kinetic part of the stress tensor defined in \cite{artoni2015average} except that we subtract the instantaneous particle-wise mean velocity instead of the time averaged one, which can eliminate the temporal velocity fluctuation of the whole layer. For example, if a box of particles were very tightly packed and its boundaries given a rigid-body oscillation in time, grains inside would still appear locked in their cage of neighbors. The temperature definition we use ensures this rigid motion is excluded from the temperature calculation, which is important as it should not influence the apparent rheology.

%%%%%%%%%%%%%%%%%%%%%%%%%%%%%%%%%%%%%%%%%%%%%%%%%%%%%%%%%%%%%%%%%%%%
\subsection*{Verification of the averaging method}
To verify the averaging method described in the main text, we compare the average stress field with the solution of the equilibrium equations in steady state:
\[
\frac{\partial \sigma_{ij}}{\partial x_j}+\rho_s \Phi G_i=0,
\]
for $i,j=\{1,2,3\}$ where $\boldsymbol{\sigma}$ is the Cauchy stress  and $\mathbf{{G}}$ is the body force (gravity). Choosing the $\hat{x}, \hat{y},\hat{z}$ basis shown in the main text, the side walls are periodic boundary conditions and we have $\frac{\partial }{\partial x}=\frac{\partial }{\partial y}=0$.  Due to the symmetry, certain stress components become determined solely from equilibrium --- i.e. are statically determinate --- which gives a means to test the validity of the stress averaging method.

\begin{figure}
    \begin{subfigure}{0.05\textwidth}
        \caption{}
        \label{fig1: 1a}
    \end{subfigure}
    \begin{subfigure}{0.37\textwidth}
        \includegraphics[width=\textwidth]{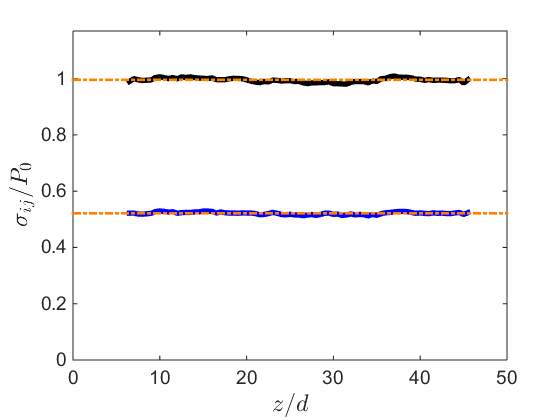}
    \end{subfigure}
    \hfill\\
    \begin{subfigure}{0.05\textwidth}
        \caption{}
        \label{fig1: 1b}
    \end{subfigure}
    \begin{subfigure}{0.37\textwidth}
        \includegraphics[width=\textwidth]{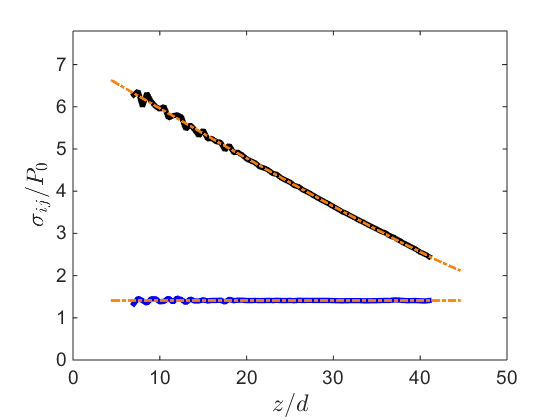}
    \end{subfigure}
    \hfill\\
    \begin{subfigure}{0.05\textwidth}
        \caption{}
        \label{fig1: 1c}
    \end{subfigure}
    \begin{subfigure}{0.37\textwidth}
        \includegraphics[width=\textwidth]{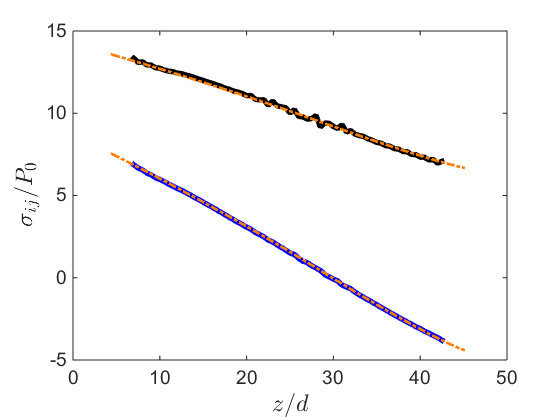}
    \end{subfigure}
    \captionsetup{justification=raggedright,
singlelinecheck=false
}
    \caption{Verification of the averaged stress, $\sigma_{zz}$ and $\sigma_{xz}$ in the three kinds of configurations tested are plotted against the height. (a) The planar shear flow ($P_f=1$, $V_f=9$); (b) the planar shear flow with gravity ($P_f=2$, $V_f=1$); (c) the inclined chute flow ($\theta=60^\circ$). The black solid lines show $-\sigma_{zz}$, the blue solid lines show $\sigma_{xz}$, the orange dash dot lines show the predicted $-\sigma_{zz}$ in (a) and (b), and the predicted $-\sigma_{zz}$ in (c) and $\sigma_{xz}$ in (a), (b), (c) up to a constant.
    }
    \label{fig1:stress test}
\end{figure}

In homogeneous planar shear flow, the stress components $\sigma_{zz}$ and $\sigma_{xz}$ throughout should be the same as the confining wall stress components. As shown in FIG. \ref{fig1: 1a}, $\sigma_{zz}=-P_{wall}$ to within $2\%$, not only showing that the averaging method is consistent with the equilibrium equations, but also meaning the feedback control method of applying top-wall pressure is capable of applying the target pressure. As shown in FIG. \ref{fig1: 1a}, the shear component $\sigma_{xz}$ is also constant in $z$, to within $2\%$.
In the cases with nonzero gravity, equilibrium gives $\partial \sigma_{zz}/\partial z=-\rho_s\Phi(z) G_z$ and $d \sigma_{xz}/d z=\partial \sigma_{xz}/\partial z=-\rho_s\Phi(z) G_x$ given $G_y=0$, which can be examined in the configurations of the planar shear flow with gravity (FIG. \ref{fig1: 1b})  and inclined chute flow (FIG. \ref{fig1: 1c}).
In the chute flow case, since the confining pressure and shear stresses at the top and bottom are not specified beforehand, solutions can be determined analytically only up to a constant.
Since $\Phi$ is in fact varying with height over the domain, we integrate $\partial \sigma_{zz}/\partial z$ and $\partial \sigma_{xz}/\partial z$ with respect to $z$
using the averaged layer-wise $\Phi(z)$ to calculate the analytical stresses.
The slow flowing regions whose strain over the sampling time is small, show expectedly stronger fluctuations about the average ($z<29d$ in FIG. \ref{fig1: 1b} and $23d<z<33d$ in FIG. \ref{fig1: 1c}).
The agreement of the averaged stress components and the values given through equilibrium in FIG. \ref{fig1:stress test} support the definitions of average packing fraction and stress used in this work. 

%%%%%%%%%%%%%%%%%%%%%%%%%%%%%%%%%%%%%%%%%%%%%%%%%%%%%%
\subsection*{Comparison with kinetic theory}
Here we make a comparison with granular kinetic theory, which can also lead to the form $gd/\delta v=F_3(\Phi)$ as mentioned in the main text, where $F_3$ is some function of $\Phi$. Regarding kinetic theory, in Lun et al.'s work \cite{lun1984kinetic}, $P$ and the viscosity $\eta$ in a granular gas are functions of $\Phi$ and granular temperature $T$:
\begin{align*}
    &P(\Phi,T)=\rho F_1(\Phi)T &\text{and} 
    &&\eta(\Phi,T)=\rho dF_2(\Phi)\sqrt{T},
\end{align*}
where $\rho$ is the density $F_1(\Phi)$ and $F_2(\Phi)$ are functions of packing fraction $\Phi$.
In Haff's work \cite{haff1983grain}, the dependency of $P$ and $\eta$ on the velocity fluctuation $\delta v$ and average separation of neighbouring grains $s$, which is a function of $\Phi$, is:
\begin{align*}
&P=t d\rho \frac{\delta v^2}{s} &\text{and} 
    &&\eta(\Phi,T)=q d^2\rho \frac{\delta v}{s},
\end{align*}
where $t$ and $q$ are dimensionless constants. Either of the theories above would give a form $gd/\delta v=F_3(\Phi)$, though neither of them predict $F_3(\Phi)$ the same as the fitted $F(\Phi)$ in the main text. 

\begin{figure}[t]
    \begin{subfigure}{0.05\textwidth}
        \caption{}
        \label{fig3: comparison(a)}
    \end{subfigure}
    \begin{subfigure}{0.37\textwidth}
        \includegraphics[width=\textwidth]{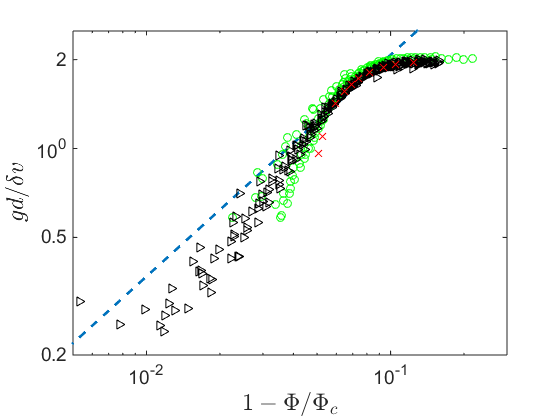}
    \end{subfigure}
    \hfill\\
    \begin{subfigure}{0.05\textwidth}
        \caption{}
        \label{fig2:test other function}
    \end{subfigure}
    \begin{subfigure}{0.37\textwidth}
        \includegraphics[width=\textwidth]{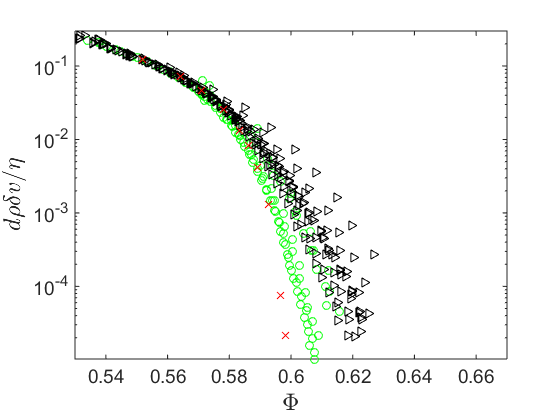}
    \end{subfigure}
    \hfill\\
    \begin{subfigure}{0.05\textwidth}
        \caption{}
        \label{fig3: comparison(b)}
    \end{subfigure}
    \begin{subfigure}{0.37\textwidth}
        \includegraphics[width=\textwidth]{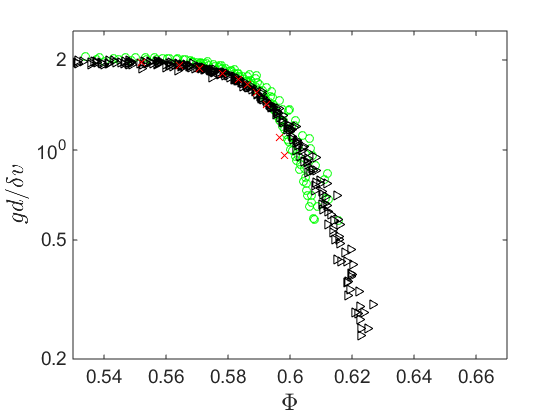}
    \end{subfigure}
    \captionsetup{justification=raggedright,
singlelinecheck=false
}
    \caption{Comparison with the locally Newtonian continuum model using DEM results in multiple configurations. (a) Normalized granular fluidity plotted against packing fraction, dashed line for the slope (0.75) proposed in \cite{losert2000particle}; (b) the viscosity form from the locally Newtonian model; (c) the scaling law of granular fluidity hypothesised in the main text. Data points from all chute flow tests ($\rhd$), planar shear with gravity tests ($\circ$) and homogeneous planar shear flow tests ($\times$).}
    \label{fig3:comparison with kinetic theory}
\end{figure}

Related to kinetic theory, Losert et al.'s locally Newtonian continuum model \cite{losert2000particle} proposes
\begin{align*}
    &P(\rho,T)=\rho T f(\rho) &\text{and} 
    &&\eta(\rho,T)=\frac{\eta_0(\rho)P(\rho,T)}{\rho_cd^2T^{1/2}},
\end{align*}
where $\rho_c$ is the density at random close packing, and $\eta_0$ is a dimensionless number proposed to be a function of packing fraction to account for the divergence of viscosity near random close packing. From the latter formula, by using $g=\dot{\gamma}/\mu$, we can directly get
\begin{align*}
    \frac{gd}{\delta v}
    =\frac{Pd}{\eta\delta v}
    =\frac{\rho_c d^3}{\eta_0}.
\end{align*}
It is supposed in \cite{losert2000particle} that
$\eta_0\sim (1-\rho/\rho_c)^{-0.75}=(1-\Phi/\Phi_c)^{-0.75}$, where $\Phi_c=0.63$ is the packing fraction of random close packing. This implies $gd/\delta v\sim (1-\Phi/\Phi_c)^{0.75}$,
which is tested in FIG. \ref{fig2:test other function}. Only some of the data can be fitted using such a power law. Substituting the pressure with its proposed density and temperature dependent formula, the locally Newtonian continuum model predicts:
\[
\frac{gd}{\delta v}
=\frac{\dot{\gamma}d}{\mu\delta v}
=\frac{\dot{\gamma}dP}{\tau\delta v}
=\frac{dP}{\eta\delta v}
=\frac{d\rho T f(\rho)}{\eta\delta v}
=\frac{d\rho_s \delta v}{\eta}f(\Phi)\Phi.
\]
This would mean, algebraically, that $d\rho_s \delta v/\eta$ should be a single function of packing fraction $\Phi$. Such a deduction is tested in FIG. \ref{fig3: comparison(a)} using the same DEM results in multiple configurations.
We notice that when the packing fraction is less than 0.57, the data points lay on a single curve, however, as the packing fraction increases, different cases have different curves. Interestingly, the normalized fluidity $gd/\delta v$ proposed in the main text and shown in FIG. \ref{fig3: comparison(b)} collapses the same data more strongly, which suggests the kinetic pressure formula used in the former test may be less reliable in dense, non-uniform flow fields. This point is discussed in more depth next.

%%%%%%%%%%%%%%%%%%%%%%%%%%%%%%%%%%%%%%%%%%%%%%%%%%%%%%
\subsection*{Possible simplifications of $g=g(\delta v, \Phi)$}
An equation of state relating the pressure, volume and temperature (PVT relationship) is a common assumption in fluids. In the hypothesized formula for $g$, $\Phi$ corresponds to ``V'' and $\delta v$ corresponds to ``T''. In the following, we discuss possible relations between $P$, $\Phi$ and $\delta v$ to see if one of the independent variables in the hypothesis $g=g(\delta v, \Phi)$ can be replaced with $P$, thereby reducing the number of physical kinematic variables introduced to one. Alternatively, it is also natural to ask if one of the independent variables tends to be `relatively constant' compared to the other, such that the profile of $g$ can be discerned largely from \emph{either} $\delta v$ or $\Phi$.  We consider these possibilities next.

\begin{figure}[t]
    \begin{subfigure}{0.05\textwidth}
        \caption{}
        \label{fig5: gn-dvn}
    \end{subfigure}
    \begin{subfigure}{0.37\textwidth}
        \includegraphics[width=\textwidth]{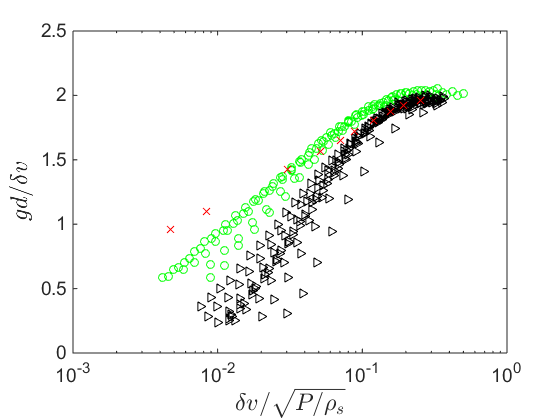}
    \end{subfigure}
    \hfill\\
    \begin{subfigure}{0.05\textwidth}
        \caption{}
        \label{fig5:gnp-Phi}
    \end{subfigure}
    \begin{subfigure}{0.37\textwidth}
        \includegraphics[width=\textwidth]{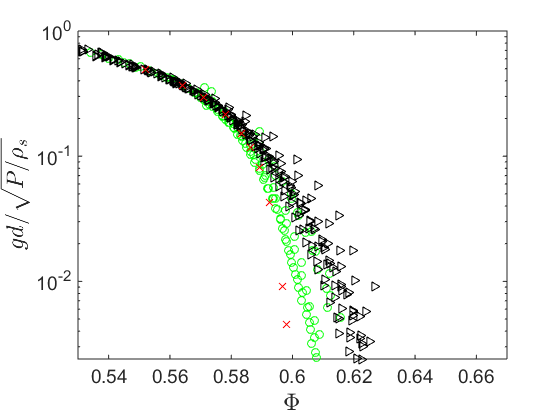}
    \end{subfigure}
    \hfill   \\
    \begin{subfigure}{0.05\textwidth}
        \caption{}
        \label{fig5: Pn-Phi}
    \end{subfigure}
    \begin{subfigure}{0.37\textwidth}
        \includegraphics[width=\textwidth]{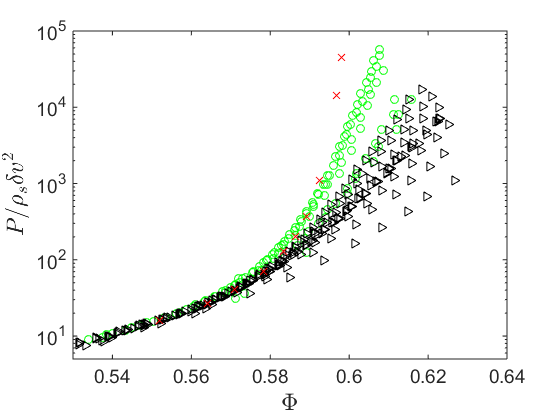}
    \end{subfigure}
    \captionsetup{justification=raggedright,
singlelinecheck=false
}
    \caption{Test with one of the independent variables of $g=g(\delta v, \Phi)$ replaced with $P$:
    (a) test $g=g(\delta v,P)$
    (b) test $g=g(\Phi,P)$.
    Data points from all chute flow tests ($\rhd$), planar shear with gravity tests ($\circ$) and homogeneous planar shear flow tests ($\times$).(c) Normalized pressure plotted against packing fraction. Data points from all chute flow tests ($\rhd$), planar shear with gravity tests ($\circ$) and homogeneous planar shear flow tests ($\times$).}
    \label{fig5:comparison with kinetic theory}
\end{figure}

\begin{enumerate}

\item \textit{Is it possible $g=g(\delta v, P)$ or $g=g(\Phi, P)$?}

If we are only allowed to use $\delta v$ and $P$ as the independent variables, then there are two dimensionless numbers: $gd/\delta v$ and $\delta v/\sqrt{P/\rho_s}$, where $\rho_s$ is the solid density of the grains. FIG. \ref{fig5: gn-dvn} shows that fluidity $g$ is not generally well described solely by $\delta v$ and $P$.

Considering $\Phi$ and $P$ as the independent variables, we can get two dimensionless numbers: $gd/\sqrt{P/\rho_s}$ and $\Phi$. FIG. \ref{fig5:gnp-Phi} shows that fluidity $g$ is not generally well described solely by $\Phi$ and $P$.

These suggest a non-unique relationship between $P$, $\Phi$ and $\delta v$ in dense inhomogeneous flows. In the absence of any dimensional constant, using $P$, $\Phi$ and $\delta v$, we can get two dimensionless numbers: $P/\rho_s\delta v^2$ and $\Phi$. Existence of a PVT equation of state would thus imply a one-to-one relation between $P/\rho_s\delta v^2$ and $\Phi$. As plotted in FIG. \ref{fig5: Pn-Phi} such an equation of state does not collapse out of the data from our many tests in inhomogeneous flows. In particular, the spread is most evident for high packing fractions where a larger role for cooperativity exists, conceivably influencing the interrelationship between these variables.  It is certainly the case that for static, dense packings, the density and pressure need not related uniquely for hard particles.

\item \textit{Is $g$ typically sensitive both to $\delta v$ and to $\Phi$, or is one variable dominant?}

 As shown in the main text, the collapse of $gd/\delta v$ onto $F(\Phi)$ spans an order of magnitude.  Hence, at hypothetically fixed velocity fluctuations, the fluidity would still be nontrivially influenced by $\Phi$ in typical flows.  The fluidity is proportional to $\delta v$ at fixed $\Phi$, so nontrivial variations in $\delta v$ would cause nontrivial variations in $g$.  Unlike equilibrium materials where internal temperature rapidly equilibrates with an external bath,  all fluctuations in granular flow are generated from plastic flow itself, absent wall vibrations.  This can produce largely inhomogeneous spatial fields for $\delta v$, varying one or more orders of magnitude in typical flows \cite{midi2004dense,koval2009annular}.  Hence, $\delta v$ cannot be treated as a relatively constant field for the purposes of computing granular fluidity in an arbitrary flow. 
 
 As a direct example, the depth profiles of velocity in inclined plane flow geometries having the same inclination angle but different heights of the granular layers  differ from each other \cite{silbert2003granular}, whereas the packing fraction and $\mu$ profiles are quite homogeneous and almost at the same value through all cases. Because $g=\dot{\gamma}/\mu=\delta v F(\Phi)/d$, the observed variations in strain-rate imply $\Phi$ is not adequate on its own to describe the granular fluidity in inclined flows.
 The inclined flow cases show that the effect of $\delta v$ is not negligible and $g=g(\delta v, \Phi)$ is generally not able to be simplified further.
\end{enumerate}

\end{document}